\documentclass[sigconf, nonacm]{acmart}
\AtBeginDocument{%
  }

\begin{document}
\setlength{\textfloatsep}{8pt}
\pagestyle{plain}

\title{MARLIN: Multi-Agent Game-Theoretic Reinforcement Learning for Sustainable LLM Inference in Cloud Datacenters}

\author{Hayden Moore}
\orcid{0009-0005-7693-703X}
\email{hayden.moore@colostate.edu}
\affiliation{%
  \institution{Colorado State University}
  \city{Fort Collins}
  \state{CO}
  \country{USA}
}

\author{Sirui Qi}
\email{alex.qi@colostate.edu}
\affiliation{%
  \institution{Colorado State University}
  \city{Fort Collins}
  \state{CO}
  \country{USA}
}

\author{Dejan Milojicic}
\email{dejan.milojicic@hpe.com}
\affiliation{%
  \institution{Hewlett Packard Labs}
  \city{Milpitas}
  \state{CA}
  \country{USA}
}

\author{Cullen Bash}
\email{cullen.bash@hpe.com}
\affiliation{%
  \institution{Hewlett Packard Labs}
  \city{Milpitas}
  \state{CA}
  \country{USA}
}

\author{Sudeep Pasricha}
\email{sudeep@colostate.edu}
\affiliation{%
  \institution{Colorado State University}
  \city{Fort Collins}
  \state{CO}
  \country{USA}
}

%\renewcommand{\shortauthors}{Moore et al.}

\begin{abstract}
Large Language Models (LLMs) have become increasingly prevalent in cloud-based platforms, propelled by the introduction of AI-based consumer and enterprise services. LLM inference requests in particular account for up to 90\% of total LLM lifecycle energy use, dwarfing training energy costs. The rising volume of LLM inference requests is increasing environmental footprints, particularly carbon emissions and water consumption. To improve sustainability for LLM inference serving in cloud datacenter environments, we propose a novel multi-agent game-theoretic reinforcement learning framework called MARLIN to co-optimize time-to-first token (TTFT), carbon emissions, water usage, and energy costs associated with LLM inference. MARLIN demonstrates a reduction of at least 18\% in TTFT, 33\% in carbon emissions, 43\% in water usage, and 11\% in energy costs compared to state-of-the-art LLM inference management frameworks.
\end{abstract}




%%\received{2 April 2026}
%%\received[accepted]{3 May 2026}

\maketitle

\section{Introduction}
Large Language Models (LLMs) have experienced explosive growth in recent years, with enterprise adoption expected to exceed 80\% in 2026 and global generative AI spending projected to reach \$644 billion \cite{ariffud2026llmstatistics}. This growth is reflected in usage data from large AI companies, with OpenAI reporting over 2.5 billion prompts per day from users across its ChatGPT deployments in 2025 \cite{shubham2026chatgpt}. With this rapid growth in LLM adoption, the associated energy usage in cloud datacenters hosting these LLMs is also rapidly increasing, with an estimated 25-33\% compound annual growth rate \cite{li2024unseenaidisruptionspower}.

The rapid increase in LLM-associated energy consumption in cloud datacenters is accelerating the already high datacenter energy footprint. In 2024, U.S. datacenters consumed 183 terawatt-hours (TWh) of electricity, which is 4\% of the total electricity generated in the country. By 2030, this figure is projected to grow by 133\% to 426 TWh \cite{leppert2026energyuse}. This increase is expected to be driven by demand for both LLM training and serving user inference requests. In the past, the training phase was the focus for improving the sustainability of LLMs. However, recent data suggest that inference dominates LLM lifecycle energy usage, accounting for 90\% of the overall energy used in this stage \cite{jegham2025hungryaibenchmarkingenergy}. Hence, it is increasingly important to focus on reducing energy costs and improving the sustainability of LLMs by reducing the resources required to serve LLM inference requests.

We target two important sustainability metrics: operational carbon emissions and water usage. Carbon emissions have been a long-term concern associated with high energy usage. LLM inference is currently scaling quickly and is already becoming a concern for carbon emissions. As an example, the yearly carbon emissions from the inference serving of just the GPT-4o model already surpass those of 30,000 cars in a year \cite{jegham2025hungryaibenchmarkingenergy}. As more competing LLMs launch and infrastructure continues to scale, emissions will only climb. 

The high water usage associated with cooling LLM-serving datacenters and during electricity generation is also becoming a concern. Cooling accounts for an estimated 29\% of datacenter water usage, with electricity generation contributing 71\%, so both factors must be considered \cite{xiao2025environmental}. For GPT-4o inference only, its annual water consumption is up to 1,579,680 kL, enough to fill more than 500 Olympic-sized pools \cite{jegham2025hungryaibenchmarkingenergy}. Importantly, this consumption refers to evaporated freshwater permanently removed from local ecosystems, straining the infrastructure of large cities, especially since an estimated 30\% of cities with over 1 million people worldwide already are facing water scarcity \cite{he2021future}. 

Clearly, the current path to scaling LLM inference is unsustainable. There is a critical need for serving LLM inference sustainably without sacrificing performance. To achieve this goal, we propose a novel framework called MARLIN, to optimize carbon emissions, water usage, and energy costs associated with LLM inference while ensuring that performance (time-to-first-token (TTFT)) remains acceptable for each incoming request. The novel contributions of our work are:
\begin{itemize}
\item We develop a novel multi-agent reinforcement learning framework called MARLIN that utilizes a game-theoretic approach to balance among competing agents.
\item We formulate a multi-objective LLM inference scheduling problem that accurately models the TTFT, carbon emissions, water usage, and energy costs of LLM serving across geo-distributed cloud datacenters.
\item We perform comprehensive ablation analysis and comparison studies with several state-of-the-art LLM inference scheduling frameworks to demonstrate the flexibility and promise of the MARLIN framework.
\end{itemize}

\section{Related Works}

Many prior efforts have addressed workload scheduling in cloud datacenters. The primary focus of optimizing workload scheduling in these works has been to improve performance and reduce overall energy consumption. For instance, Hogade et al. proposed a game-theoretic framework that incorporates network transfer costs to minimize datacenter energy costs \cite{9445675}. Chen et al. used ant colony optimization to consolidate VMs, reducing both power consumption and thermal hotspots \cite{CHEN2023578}.

Recent efforts have shifted towards optimizing sustainability metrics, particularly carbon emissions and water usage. To address the rising datacenter carbon emissions, Qi et al. developed a dual-objective evolutionary optimizer that improved the carbon emissions and throughput of serverless function execution across geo-distributed datacenters \cite{10765835}. Beena et al. proposed a greedy scheduling heuristic that worked across many distributed datacenters and used an LSTM model for carbon emission prediction \cite{10971939}. Qi et al. minimized carbon emissions, wastewater, and costs using a hybrid evolutionary-boosting algorithm \cite{11433669}. These prior approaches do not address sustainable LLM inference scheduling.

Most existing LLM inference schedulers prioritize datacenter performance metrics. For example, Mei et al. used mixed-integer linear programming to maximize data flow across distributed GPUs to maximize LLM inference throughput \cite{10.1145/3669940.3707215}. Patel et al. divided LLM inference into distinct phases, enabling pipeline parallelism to improve throughput \cite{10609649}. Yang et al. developed a framework that leveraged edge and cloud servers using constraint satisfaction and upper confidence bounds, thereby improving throughput and costs \cite{yang2024perllmpersonalizedinferencescheduling}. Recent work is starting to focus on sustainability in LLM inference scheduling. Moore et al. combined a genetic algorithm and machine learning to generate Pareto-optimal solutions for LLM inference serving \cite{10.1145/3716368.3735301}. However, the approach lacks scalability and has a slow convergence speed.

MARLIN addresses these limitations with a multi-agent game-theoretic RL approach that improves convergence speed, preserves optimization quality, and balances sustainability with TTFT performance.

\section{Modeling Overview}
Our framework is designed to map incoming LLM inference requests to server nodes across geo-distributed datacenters to leverage natural geographic and temporal variations. We comprehensively model the LLM inference workload and cloud datacenter characteristics across geographic locations, capturing energy costs, carbon emissions, water usage, and data transmission latency for each request.
\begin{figure}
  \centering
  \includegraphics[width=\linewidth]{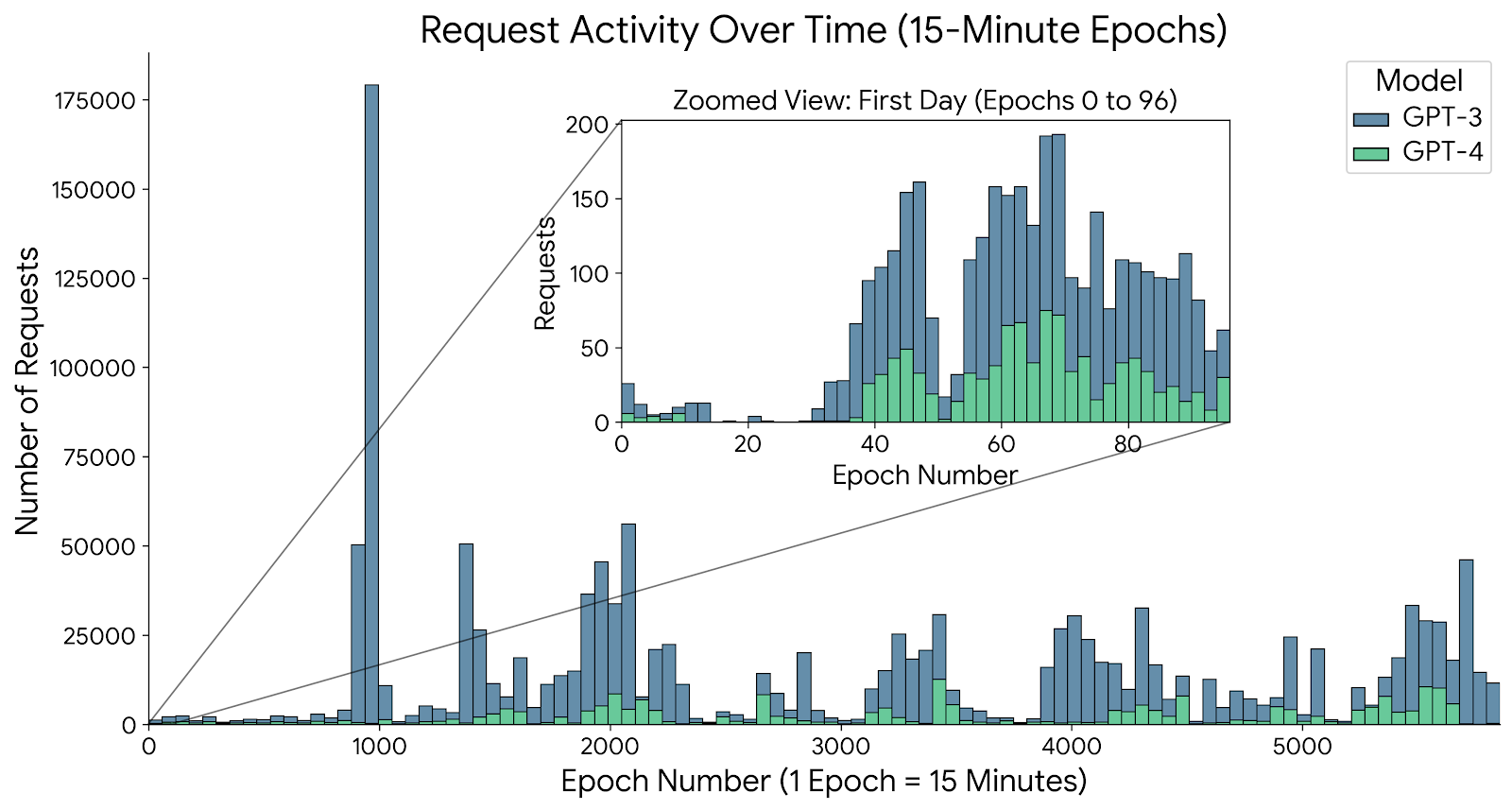}
  \caption{The number of individual LLM requests in each epoch (15 minutes) over two weeks \cite{10.1145/3711896.3737413}}
  \Description{Stacked bar histogram titled ``Request Activity Over Time (15-Minute Epochs).'' The horizontal axis shows epoch number from 0 to approximately 5600, where each epoch is 15 minutes, spanning two weeks. The vertical axis shows number of requests from 0 to 175000. Bars are stacked for two models: GPT-3 in blue and GPT-4 in green. A dominant spike near epoch 1000 reaches approximately 175000 requests. Activity across the full trace is bursty, with intermittent peaks of 10000 to 55000 requests separated by near-zero valleys. An inset zooms into the first day (epochs 0 to 96), showing GPT-3 and GPT-4 requests rising from near zero to a peak of roughly 190 around epoch 65, then declining, with GPT-4 comprising a notable fraction of traffic from epoch 40 onward.}
  \label{fig1:workload_analysis}
\end{figure}
\subsection{LLM Workload Model}

Our workload model uses a real-world trace \cite{10.1145/3711896.3737413} from a ChatGPT service on Azure cloud datacenters. The dataset covers two weeks' worth of requests for GPT-3 and GPT-4. We aggregate requests by epoch (15-minute intervals) and plot their frequencies in Fig. \ref{fig1:workload_analysis}. The distribution of LLM inference requests across epochs is quite diverse, highlighting the challenge of scheduling them across cloud datacenters. We created a custom workload model that used the observed arrival patterns from the real cloud trace and paired them with execution models for two contemporary LLMs, Llama-7B and Llama-70B. Execution profiles for these two LLMs were available across multiple GPUs \cite{touvron2023llama2openfoundation}.

Real-time LLM serving requires meeting strict memory constraints. Memory constraints require that each request $i$’s memory footprint $MF_i$ not exceed a server node $n$’s maximum available memory. Requests share the underlying model’s weights when possible, to reduce memory overhead. We account for shared weights for a model $\upsilon$’s memory footprint $MF_{\upsilon}$, and each request’s growing KV cache $KV_{(\upsilon,i,\tau)}$ as a function of the number of tokens $\tau$. KV cache growth is proportional to the model size. The total server node memory usage $M_{(n,e)}$ computed for all tokens $T$ at epoch $e$ is:
\begin{equation}
M_{n,e}=(\sum_{\upsilon}^{\Upsilon}MF_{\upsilon})+(\sum_i^{I_e}\sum_{\tau}^{T}KV_{\upsilon, i, \tau})
\end{equation}
\noindent where $\Upsilon$ is the total number of all active LLMs and $I_e$ is the total number of requests at epoch $e$. If this memory usage exceeds the node $n$’s maximum available memory $M_{tot,n}$, it results in a queuing of requests until prior requests release their KV cache.
Real-time LLM serving also requires meeting strict latency constraints. As many LLM applications (e.g., chatbots) require real-time interactivity, minimizing latency $LA_i$ for each request is critical. Latency in this case has three components: LLM weight loading time, network latency, and computation time.

Weight loading time $LA_{load,i}$  depends on whether the model is present on the GPU during execution. We assume each model is present on the node locally in storage. However, the model cannot be used until it is loaded onto the onboard memory. The worst-case loading time depends on the slowest memory bandwidth $BW_n$ in moving the model to the onboard memory. Therefore, the model weight loading time can be expressed as: $LA_{load,i}=MF_\upsilon/BW_n$. To reduce loading time for additional requests, a node keeps the model present in onboard memory.

The network latency $LA_{(net,i)}$ consists of moving the user request to a datacenter. The transfer latency of the user request to the datacenter is dependent on the distance $dist$ and transmission media characterized by $\lambda_{media}$ \cite{8098560}, which is the propagation latency per unit of distance (e.g., milliseconds per kilometer).  Requests may also need to be transmitted between datacenters via an inter-datacenter network, which depends on hop count $R_{source,dest}$ as well as the latency of processing each hop $\sigma_{hop}$.  The network latency $LA_{net,i}$ can be expressed as:
\begin{equation}
LA_{net,i}= (dist \times \lambda_{media})+(R_{source,dest} \times \sigma_{hop})
\end{equation}
At the scheduled server node $n$, computation proceeds until the first output token $\tau$ is generated. We assume profiled execution time $LA_{(tot,exec,i)}$ scales linearly with output token count $T_i$. We can safely make this assumption due to batched LLM inference being memory bound on GPUs \cite{11120553}. This bounds the throughput giving the execution time a linear relationship.

To calculate the overall TTFT, all latency components must be combined, with network latency counted twice in the final calculation to account for both sending and receiving the request. Therefore, the total request TTFT across all requests $I$ during an epoch $e$ can be estimated as:
\begin{equation}
LA_{tot,e}=\sum_i^{I_e}(LA_{load,i}+2\times LA_{net,i}+LA_{tot,exec,i}/T_i )
\end{equation}
\subsection{Datacenter Model}

Each geo-distributed datacenter $d$ comprises $N_d$ nodes at a location. All datacenters that are present in the geo-distributed network are defined by $D$, where each datacenter is present at a unique location. Nodes are organized in standard hot/cold aisle layouts with conventional air cooling \cite{robert2000alternating}. Nodes vary in GPU types and configurations. Each node $n$ contains identical GPUs whose memory is pooled to serve incoming requests. We consider the NVIDIA A100 and H100 as the GPUs present in the nodes. The number of GPUs per node can be 2, 4, or 8, allowing each node to host models of varying complexity.
\subsection{Energy Cost Model}

Each server node $n$ can operate in preset performance states, each of which dissipates a fraction $PC_{pstate}$  of the node’s total thermal design power $TDP_n$. The information technology (IT) node energy over an epoch $e$ of duration $t_e$ is computed as:
\begin{equation}
E_{IT,n,e}=PC_{pstate}  \times TDP_n  \times t_e
\end{equation}

The total datacenter IT energy over an epoch is the sum of contributions from all its server nodes:
\begin{equation}
E_{IT,d,e}= \sum_n^{N_d}E_{IT,n,e} 
\end{equation}

Datacenter energy also includes cooling energy $E_{cool,d,e}$ and infrastructure energy $E_{infra,d,e}$. Cooling energy consists of the energy required to run the computer room air conditioning (CRAC) units. CRAC efficiency (COP) varies by datacenter, $COP_d$. Across an epoch $e$, the CRAC energy is $E_{CRAC,d,e}=E_{IT,d,e}/COP_d$. Since chiller energy usage and auxiliary equipment energy usage are approximately equivalent to $E_{CRAC,d,e}$ \cite{ZHANG2021102253}, the total datacenter cooling energy is: $E_{cool,d,e}=3\times E_{CRAC,d,e}$.

The supporting infrastructure consists of power distribution units and power supply units in the datacenter. Infrastructure energy is equal to about 13\% of the IT energy \cite{9599719}: $E_{infra,d,e}=0.13\times E_{IT,d,e}$.

The total datacenter energy usage at a datacenter $d$ during an epoch $e$ consists of the sum of components:
\begin{equation}
E_{tot,d,e}=E_{IT,d,e}+E_{cool,d,e}+E_{infra,d,e}
\end{equation}

However, the cost of using that energy will vary depending on the location of the datacenter $d$ and the time-of-day when the epoch $e$ occurs, which determines the time-of-use pricing $TOU_{d,e}$ set by the utility provider. Therefore, the total cost of energy usage across all the datacenters for epoch $e$ is:
\begin{equation}
Cost_{tot,e}=\sum_d^D(E_{tot,d,e} \times TOU_{d,e})                          
\end{equation}
\subsection{Water Usage Model}

Water is used in two ways in a datacenter: for direct cooling and for energy generation. The two ways can be further divided into three sub-components, which are location-specific: evaporative $G_{E,d,e}$, blowdown $G_{blow,d,e}$, and grid-based $G_{grid,d,e}$. Evaporative water loss is water that exits through the cooling towers, which is proportional to the cooling load $H_{cool,d,e}$ and the latent heat of vaporization $J_{water}$: $G_{E,d,e}=H_{cool,d,e}/J_{water}$. CRACs require a coolant (typically water) for cooling, which must be treated when pollutant concentration reaches a threshold $\phi$ \cite{Siddik_2021}. The treatment requires the polluted water to be sent to a blowdown-to-water treatment facility: $G_{blow,d,e}=G_{E,d,e}/(1-\phi)$. Lastly, energy generation consumes water at location-dependent intensity $GI_d$, which is proportional to the underlying grid makeup. The variation between sources can be dramatic, with wind using only 0.2 L/kWh, whereas hydropower uses 67 L/kWh, most of which is lost to evaporation \cite{JIN2019109391}. This gives a water usage of $G_{grid,d,e}=E_{tot,d,e}\times GI_d$.

The total water usage in an epoch across all datacenters is:
\begin{equation}
G_{tot,e}=\sum_d^D(G_{E,d,e} +G_{blow,d,e}+G_{grid,d,e})                   
\end{equation}
\subsection{Carbon Emissions Model}

Carbon emissions arise from electricity generation. This has two components: grid-based emissions $Z_{grid,d,e}$ and water-treatment-based emissions $Z_{G,d,e}$. Grid-based emissions depend on carbon intensity $CI_{d,e}$ (kg of carbon per kWh produced), which depends on the energy sources in the underlying grid. This gives an emissions of: $Z_{grid,d,e}=CI_{d,e}  \times E_{tot,d,e}$

Wastewater processing consumes electricity from the local grid. However, potable water creation $EI_{pot,e}$ and wastewater treatment $EI_{waste,e}$ have different energy intensities \cite{11433669}. Potable water production arises from the need to replace evaporated water and process blowdown water. The potable carbon emissions are: $Z_{pot,d,e}=(G_{blow,d,e}+G_{E,d,e}) \times EI_{pot,e}$

The operation of the wastewater treatment plants is considered in terms of how they process grid-based water usage. The wastewater carbon emissions are: $Z_{waste,d,e}=G_{grid,d,e}\times EI_{waste,e}$

Water-associated carbon emissions at a datacenter is scaled by the carbon intensity:
\begin{equation}
Z_{G,d,e}=(Z_{pot,d,e}+Z_{waste,d,e})\times CI_{d,e}
\end{equation}

The total carbon emissions in a single epoch are the sum of all sources across all datacenters $D$:
\begin{equation}
Z_{tot,e}=\sum_d^D(Z_{grid,d,e}+Z_{G,d,e})	
\end{equation}
\section{Problem Formulation}

We assume a single cloud service provider (e.g., Meta) that manages its geo-distributed datacenters for serving incoming LLM inference requests. Requests originate remotely and thus incur associated transfer times that must also be considered. Our problem optimization objective is to minimize the weighted sum of TTFT, carbon emissions, water usage, and energy costs over $E$ epochs, given resource constraints $M_{tot,n}$ and SLA constraints (SLA):
\begin{equation}
\begin{gathered}
Min\sum_e^E(w_1 LA_{tot,e} + w_2 Z_{tot,e} + w_3 G_{tot,e} + w_4 Cost_{tot,e}) \\
s.t.        M_{n,e}  \leq M_{tot,n}       \forall n \in N \\
LA_{tot,e} \leq LA,      \forall i \in I_e \\
\sum_{j=1}^4 w_j=1,       w_j \geq 0 
\end{gathered}
\end{equation}

Our goal is to develop a geo-distributed scheduling framework that assigns each incoming LLM inference request to a datacenter while optimizing the above objective and satisfying the constraints mentioned. Once a request is assigned to a datacenter, a local load balancer algorithm distributes requests to nodes within the datacenter. All datacenters use a modified round-robin load balancer based on \cite{9377049}, which ensures that nodes within a datacenter are not asymmetrically overwhelmed with requests.
\begin{figure}
  \centering
  \includegraphics[width=\linewidth]{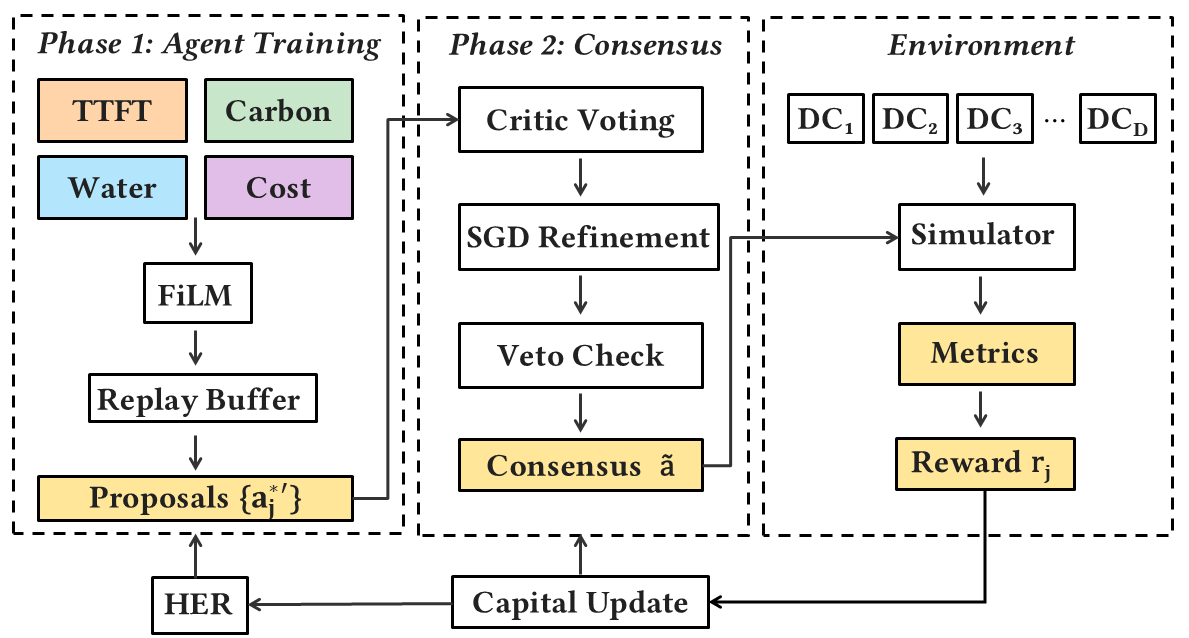}
  \caption{Overview of MARLIN framework and its two phases.}
  \Description{Block diagram of the MARLIN framework divided into three dashed-border panels. The left panel, labeled Phase 1: Agent Training, contains four colored input boxes---TTFT in orange, Carbon in green, Water in blue, and Cost in purple---feeding downward into FiLM, then Replay Buffer, then a yellow output box labeled Proposals. The center panel, labeled Phase 2: Consensus, contains a vertical sequence of boxes: Critic Voting, SGD Refinement, Veto Check, and a yellow output box labeled Consensus \tilde{a}. The right panel, labeled Environment, shows datacenters DC1 through DCD feeding into a Simulator, then a yellow Metrics box, then a yellow Reward box labeled r\_j. Arrows connect the panels: Proposals feeds into Critic Voting; SGD Refinement has a bidirectional arrow to Simulator; Reward feeds left into a Capital Update box at the bottom, which feeds into an HER box, which feeds back up into Proposals.}
  \label{fig2:Framework overview}
\end{figure}
\section{MARLIN Framework}

Fig. \ref{fig2:Framework overview} provides an overview of the MARLIN framework, which comprises agents that participate in a two-phase competitive game. To effectively service LLM requests, a scheduling plan $a$ is needed to pair requests with a datacenter to process them. In phase 1, each agent $j$ independently proposes a single scheduling plan $a_j^{*'}$ that best optimizes its single objective (e.g., TTFT, carbon, water, cost). The quality of all plans $a_j^{*'}$ in phase 1 is estimated through their respective reward value $r_j$. Each agent maintains its own reward function to score the plans. In phase 2, all $J$ agents negotiate via weighted voting to determine the final blended scheduling plan $\tilde{a}$ and agent-wise capital $[C_j]_{j=1}^J$. The final blended plan $\tilde{a}$ is determined by the agent-wise plan $a_j^{*'}$, capital threshold $\delta_j$, and utility function $Q_j$. The capital threshold $\delta_j$ determines when the agent $j$ can update the final plan according to its interest. $Q_j$ estimates the long-term importance of each proposal (vs. the short term reward in phase 1). On top of the blended plan $\tilde{a}$, each agent $j$ maintains a reserve of capital $C_j$ that is updated after each epoch in the game. Capital $C_j$ represents the influence of each agent when constituting the blended plan. We formulate phase 2 as a weighted resource-allocation game $\Gamma$ which is defined as: 
\begin{equation}
\tilde{a},[C_j]_{j=1}^J=\Gamma([a_j,\delta_j,C_j,Q_j ]_{j=1}^J)	
\end{equation}

MARLIN combines both phases with workload predictions and environmental state information to output a final blended plan $\tilde{a}$ for sustainable geo-distributed LLM inference scheduling. The following subsections describe the workload predictor, phases 1 and 2, and associated complexity analysis.
\subsection{Workload Predictor}

LLM request volume varies over time and across locations, as shown in Fig. \ref{fig1:workload_analysis}, necessitating accurate prediction to optimize resource allocations. By forecasting workloads, MARLIN can prewarm containers and avoid startup delays. We adopt and pretrain the predictor (line 1 in Algorithm 1) from \cite{10.1145/3472883.3487014}, which utilizes a regression-based approach. The predictor forecasts the number of LLM requests in the next epoch based on information from past epochs within a time window of $tw$, using an exponentially weighted moving average to estimate the incoming workload. This results in an extremely fast prediction time of roughly 100 microseconds per prediction. To evaluate the accuracy of the predictor it is compared to a baseline neural network predictor \cite{10.1145/3472883.3487014}. The regression-based predictor achieves over 90\% accuracy in empirical validation experiments across various workload intensities and volatilities, demonstrating an average accuracy improvement of 19.01\% over the baseline.
\subsection{Phase 1: Agent Training}
\begin{figure}
  \centering
  \includegraphics[width=\linewidth]{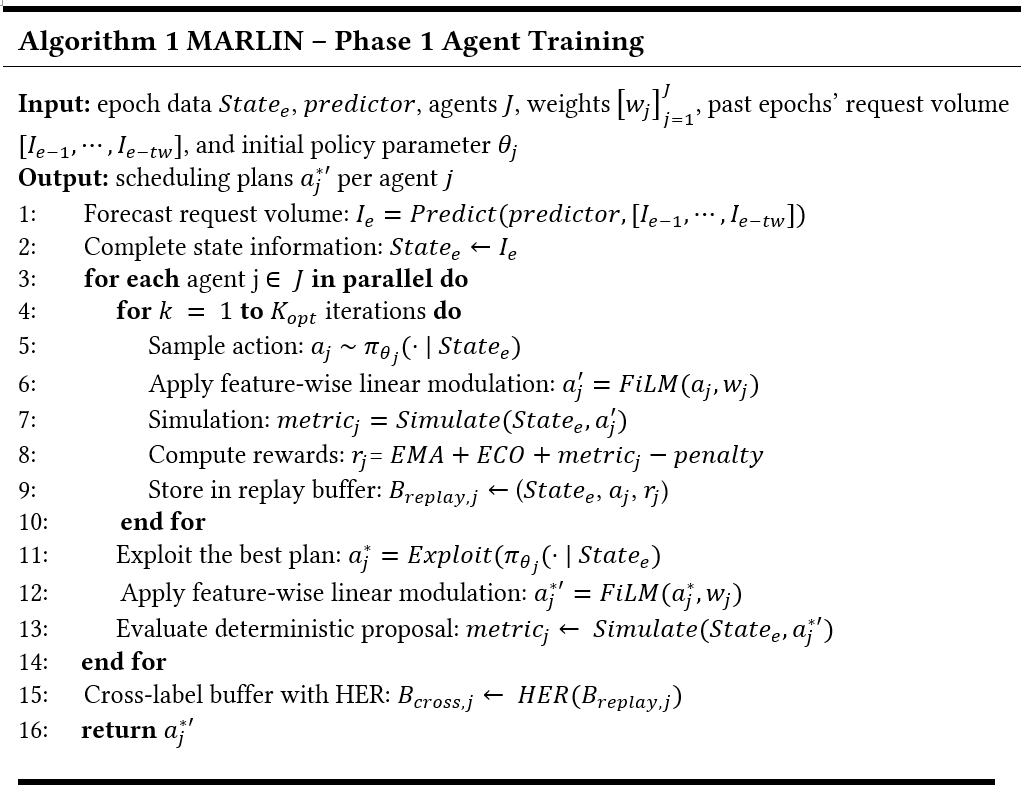}
  \Description{Pseudocode listing titled ``Algorithm 1 MARLIN -- Phase 1 Agent Training.'' Inputs are epoch data State\_e, a predictor, agents J, agent-wise weights w\_j, past epoch request volumes I\_{e-1} through I\_{e-tw}, and initial policy parameters theta\_j. Output is scheduling plans a\_j\^{}*prime per agent j. The 16-line algorithm proceeds as follows: line 1 forecasts the current epoch request volume using the predictor; line 2 adds the forecast to State\_e; lines 3 through 14 loop over all agents j in parallel, and within each agent loop over K\_opt iterations: sampling an action a\_j from the policy, applying FiLM modulation to obtain a\_j prime, simulating the modulated plan to obtain metric\_j, computing reward r\_j as EMA plus ECO plus metric\_j minus penalty, and storing the experience in replay buffer B\_replay\_j; after the inner loop, line 11 exploits the policy to obtain the best plan a\_j\^{}*, line 12 applies FiLM modulation to yield a\_j\^{}*prime, and line 13 evaluates this plan via simulation; line 15 applies HER to produce a cross-epoch buffer B\_cross\_j; line 16 returns a\_j\^{}*prime.}
  \label{fig3:Algo1}
\end{figure}

In phase 1, we use a multi-agent soft actor-critic (SAC) algorithm with feature-wise linear modulation (FiLM) layer \cite{perez2018film} and hindsight experience replay (HER) \cite{NIPS2017_453fadbd}, as shown in Fig. 2. SAC is an actor-critic variant that uses stochastic off-policy learning to improve exploration. We add a FiLM layer to the actor’s MLP for improved feature modulation. HER is a technique that modifies the replay buffer labels and enables training in such a complex environment. The details of phase 1 are presented in Algorithm 1. 

The algorithm takes initial state information for the current epoch $State_e$, pretrained predictor $predictor$, number of agents $J$, agent-wise weights $[w_j]_{j=1}^J$, the past epochs’ request volume $[I_{e-1},...,I_{e-tw}]$, and the initial policy parameter $\theta_j$ as input for agent training. After SAC training with the help of FiLM layer and HER, the algorithm outputs agent-wise plan $a_j^{*'}$.

The $predictor$ forecasts the current epoch’s LLM workload (lines 1-2). This becomes part of the state information $State_e$ of the current epoch, which also includes information about each datacenter’s node availability, carbon intensity, water intensity, and energy pricing. All agents are trained in parallel and iterate over $K_{opt}$ iterations to update their policy parameter and Q function (lines 3-4). In each iteration, the agent $j$ will first take an action $a_j$ as a scheduling plan and then apply feature-wise linear modulation on $a_j$ for a modulated plan $a_j^{'}$ (lines 5-6). The FiLM layer applies affine transformations across the neural network input. This integrates workload data and datacenter attributes into a single input. Based on the modulated plan $a_j^{'}$ the optimization metrics are estimated (line 7) to be $metric_j=[LA_{tot},Z_{tot},G_{tot},Cost_{tot}]_{a_j}$

We calculate the reward value $r_j$ for modulated plan $a_j^{'}$ (line 8) by considering an exponential moving average ($EMA$), $metric_j$, an ecological bonus that rewards consolidation $ECO$, and an SLA $penalty$ that punishes TTFT beyond a set point in the reward value estimation equation which is $r_j= EMA+ECO+metric_j-penalty$.

The SAC training experience is stored in the replay buffer $B_{replay,j}$ (line 9). After $K_{opt}$ iterations, agent $j$ generates the best plan $a_j^*$ by exploiting the policy parameter $\theta_j$ (line 11). We apply FiLM to the best plan $a_j^*$ to obtain the best modulated plan $a_j^{*'}$ (line 12). Based on the modulated plan $a_j^{*'}$, we estimate $metric_j$ through simulation (line 13). To balance current-epoch learning with historical insights, we maintain a cross-epoch buffer $B_{cross,j}$ for each agent $j$. The cross-epoch buffer is updated per epoch by sampling from the replay buffer $B_{replay,j}$ with HER (line 14).
\subsection{Phase 2: Competitive Proposals}
\begin{figure}
  \centering
  \includegraphics[width=\linewidth]{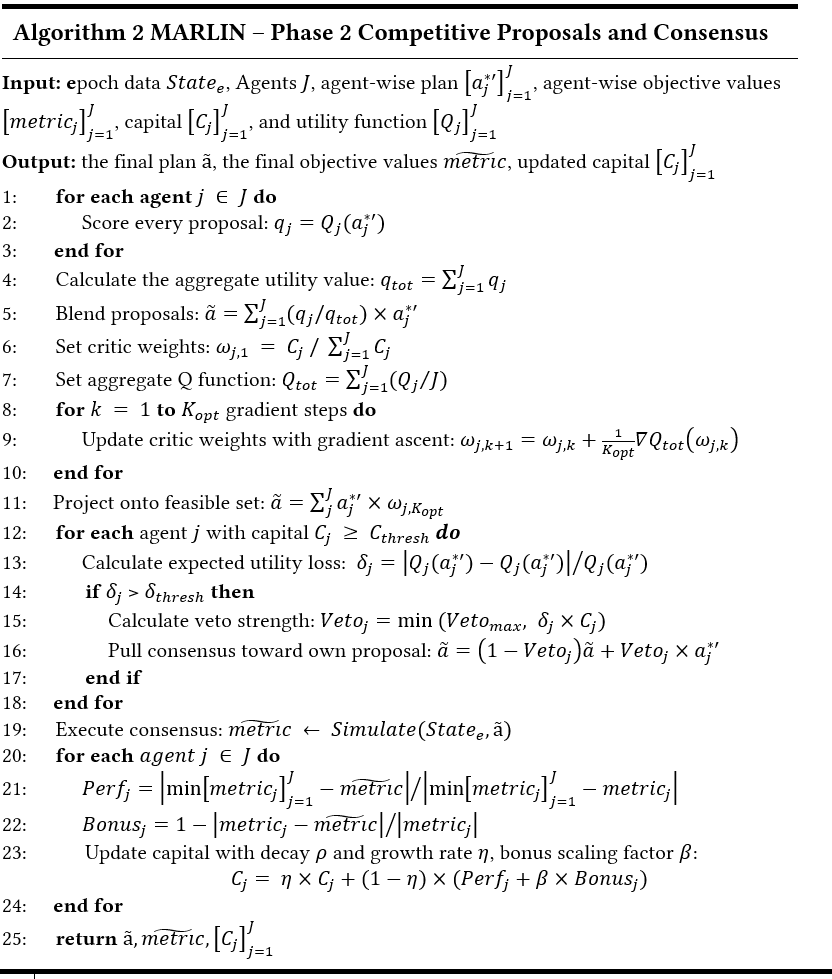}
  \Description{Pseudocode listing titled ``Algorithm 2 MARLIN -- Phase 2 Competitive Proposals and Consensus.'' Inputs are epoch data State\_e, agents J, agent-wise plans a\_j\^{}*prime, agent-wise objective values metric\_j, capital C\_j, and utility functions Q\_j. Outputs are the final plan \~{a}, final objective values metric-tilde, and updated capital C\_j. The 25-line algorithm proceeds as follows: lines 1 through 3 score each agent proposal q\_j using its utility function; line 4 computes the aggregate utility q\_tot; line 5 forms an initial blended plan as a weighted sum of proposals scaled by q\_j over q\_tot; line 6 sets initial critic weights omega\_j proportional to capital C\_j; line 7 defines an aggregate Q function Q\_tot as the mean of all Q\_j; lines 8 through 10 refine the blended plan over K\_opt gradient ascent steps on Q\_tot; line 11 projects the refined plan back onto the feasible set; lines 12 through 18 implement the veto mechanism, where each agent with sufficient capital computes utility loss delta\_j, and if delta\_j exceeds delta\_thresh the agent pulls the blended plan toward its own proposal proportional to veto strength Veto\_j; line 19 simulates the consensus plan to obtain final metrics; lines 20 through 24 update each agent's capital using performance score Perf\_j, bonus score Bonus\_j, growth rate eta, and scaling factor beta via a bounded exponential moving average; line 25 returns the final plan, metrics, and updated capitals.}
  \label{fig4:Algo2}
\end{figure}
Phase 2 focuses on combining the separate scheduling plans $a_j^{*'}$ generated from phase 1 into a single blended plan $\tilde{a}$. We employ game theory to optimize this blending process. The agent’s associated capital $C_j$ is essential in advancing its own plan in the final blended solution. Algorithm 2 presents the details of phase 2.

First, we calculate the utility value $q_j$ for each agent $j$ (lines 1-2). After that, we calculate the aggregated utility value $q_{tot}$ for all agents $J$ (line 4 in Algorithm 2). The weight of each agent’s plan in the blended plan is determined by the corresponding utility value $q_j$ scaled by the aggregated utility value $q_{tot}$ (line 5). To help formulate the blended plan, we calculate the initial critic weight $\omega_j$ based on each agent’s capital $C_j$ (line 6). Moreover, an aggregate Q function $Q_{tot}$ is formulated to estimate the long-term quality of the blended plan $\tilde{a}$ (line 7). The blended plan is refined over $K_{opt}$ iterations via stochastic gradient descent (SGD) \cite{ruder2017overviewgradientdescentoptimization} to maximize the aggregate Q function (lines 8-10). After each SGD step, a new blended plan is created via the critic weights (line 11).
\begin{figure*}
  \centering
  \includegraphics[width=\linewidth]{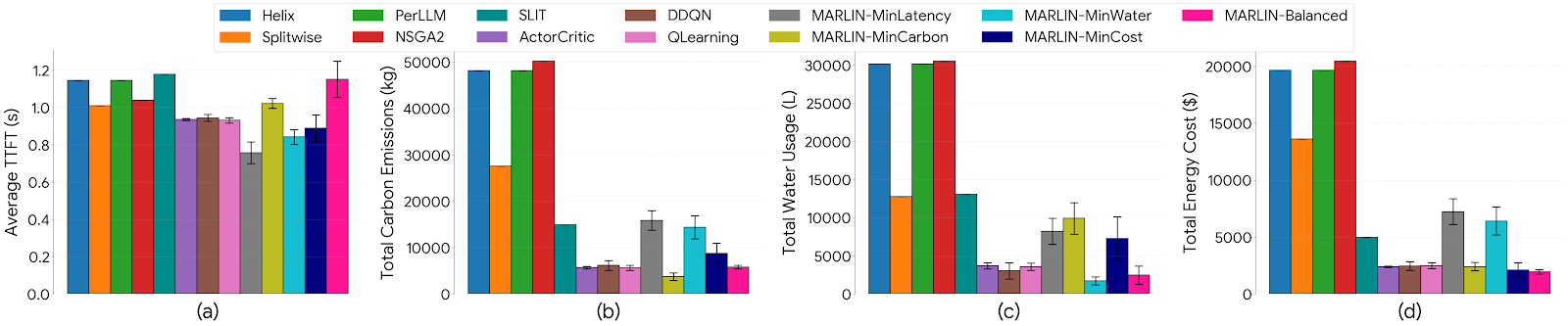}
  \caption{Comparison of the (a) TTFT, (b) carbon emissions, (c) water usage, and (d) energy costs, across frameworks.}
  \Description{Four-panel grouped bar chart comparing 13 LLM inference scheduling frameworks. The legend identifies the frameworks by color: Helix in blue, Splitwise in orange, PerLLM in green, NSGA2 in red, SLIT in teal, ActorCritic in purple, DDQN in brown, QLearning in pink, MARLIN-MinLatency in gray, MARLIN-MinCarbon in yellow-green, MARLIN-MinWater in cyan, MARLIN-MinCost in dark navy, and MARLIN-Balanced in magenta. Panel (a) shows average TTFT in seconds on a 0 to 1.2 scale. Helix, PerLLM, NSGA2, Splitwise, and SLIT cluster between 0.9 and 1.2 seconds. DDQN, QLearning, and ActorCritic fall between 0.75 and 0.95 seconds. MARLIN-MinLatency achieves approximately 0.75 seconds, MARLIN-MinCost and MARLIN-MinCarbon near 0.85 seconds, MARLIN-MinWater near 0.9 seconds, and MARLIN-Balanced near 1.15 seconds. Panel (b) shows total carbon emissions in kilograms on a 0 to 50000 scale. Helix, PerLLM, and NSGA2 reach approximately 47000 to 50000 kg. Splitwise drops to roughly 28000 kg. SLIT is near 16000 kg. The three RL baselines fall between 5000 and 17000 kg with error bars. MARLIN-MinCarbon is the lowest at roughly 3000 kg; MARLIN-MinWater, MARLIN-MinCost, and MARLIN-Balanced range from 4000 to 9000 kg. Panel (c) shows total water usage in liters on a 0 to 30000 scale. Helix, PerLLM, and NSGA2 reach approximately 30000 L. Splitwise is near 12500 L. SLIT is near 13000 L. RL baselines fall between 4000 and 9000 L. MARLIN-MinWater is the lowest near 1000 L; MARLIN-MinCarbon, MARLIN-MinCost, and MARLIN-Balanced range from 2000 to 8000 L. Panel (d) shows total energy cost in dollars on a 0 to 20000 scale. Helix, PerLLM, and NSGA2 are near 19000 to 21000 dollars. Splitwise is near 13500 dollars. SLIT and RL baselines range from 3000 to 15000 dollars. MARLIN-MinCost is the lowest near 2200 dollars; other MARLIN variants range from 2500 to 6500 dollars.}
  \label{fig3:baseline_results}
\end{figure*}

To maintain a balanced trade-off among all agents, a veto mechanism is designed based on the individual rationality (IR) theory \cite{von1947theory}. IR is a game theory approach in which a participant will engage only if their expected payoff exceeds that of not participating. This holds for the veto mechanic, as agents are not required to veto in every single epoch and will only consider using their veto under the following circumstances: if 1) their capital exceeds a threshold $C_{thresh}$ and 2) their critic network $Q_j$ estimates that the consensus provides a utility loss $\theta_j$ more than $\theta_{thresh}$ compared to its own proposal (lines 12-14). If a veto scenario is triggered, we calculate the veto strength $Veto_j$ based on utility value degradation $\theta_j$ and maximal veto strength $Veto_j$ (line 15). Based on the veto strength, we update the blended plan $\tilde{a}$ by moving the blended plan toward the agent’s plan $a_j^{*'}$ (line 16).

Once all agents are satisfied with the blended plan (no threshold $C_{thresh}$ or $\theta_{thresh}$ is violated), the optimization metric values $\widetilde{metric}$ are estimated accordingly via simulation (line 19). To pass down the knowledge we learn from the veto process, we need to update the capital $C_j$ used in the veto mechanism for the next epoch. We first calculate the performance score $Perf_j$ for each agent’s capital $C_j$ (line 21). The performance score estimates how well the current blended plan $\tilde{a}$ serves the agent’s $metric_j$. After that, the bonus score $Bonus_j$ is calculated to estimate each agent's contribution to the final blended plan (line 22).  Given the performance score $Perf_j$, bonus score $Bonus_j$, predefined growth rate $\eta$, and predefined scaling factor $\beta$, we update all the capitals using a bounded EMA approach (line 23). 

\subsection{Complexity Analysis}

The runtime and memory complexity of the framework depend on the number of datacenters $D$, policy network parameters $\theta_j$, number of agents $J$, replay buffer size $B_{replay,j}$, and exploration steps $K_{opt}$.

In phase 1, the runtime complexity is driven by exploration steps $K_{opt}$ and the policy network parameters $\theta_j$. Agents $J$ are run in parallel. Therefore, the final runtime complexity is $O(K_{opt}\times \theta_j)$. The memory complexity of phase 1 relies on the size of the replay buffers $B_{replay,j}$ as well as the policy network parameters $\theta_j$. The buffer and the policy network sizes both grow with the total agents $J$. The memory complexity of phase 1 is therefore $O(J\times \theta_j+J \times B_{replay,j} \times D)$.

As the iterations of the blended plan are performed across one plan, the runtime complexity of phase 2 is identical to phase 1 as $O(K_{opt} \times \theta_j)$. The memory complexity of phase 2 is much lower than that of phase 1. The only two memory requirements are the holding of the proposals $a_j^{*'}$ from phase 1 and the agents $J$ themselves. The proposal footprint from phase 1 is proportional to the number of datacenters $D$. Therefore, the memory complexity of phase 2 is $O(J\times \theta_j+J \times D)$.
\begin{figure}
  \centering
  \includegraphics[width=\linewidth]{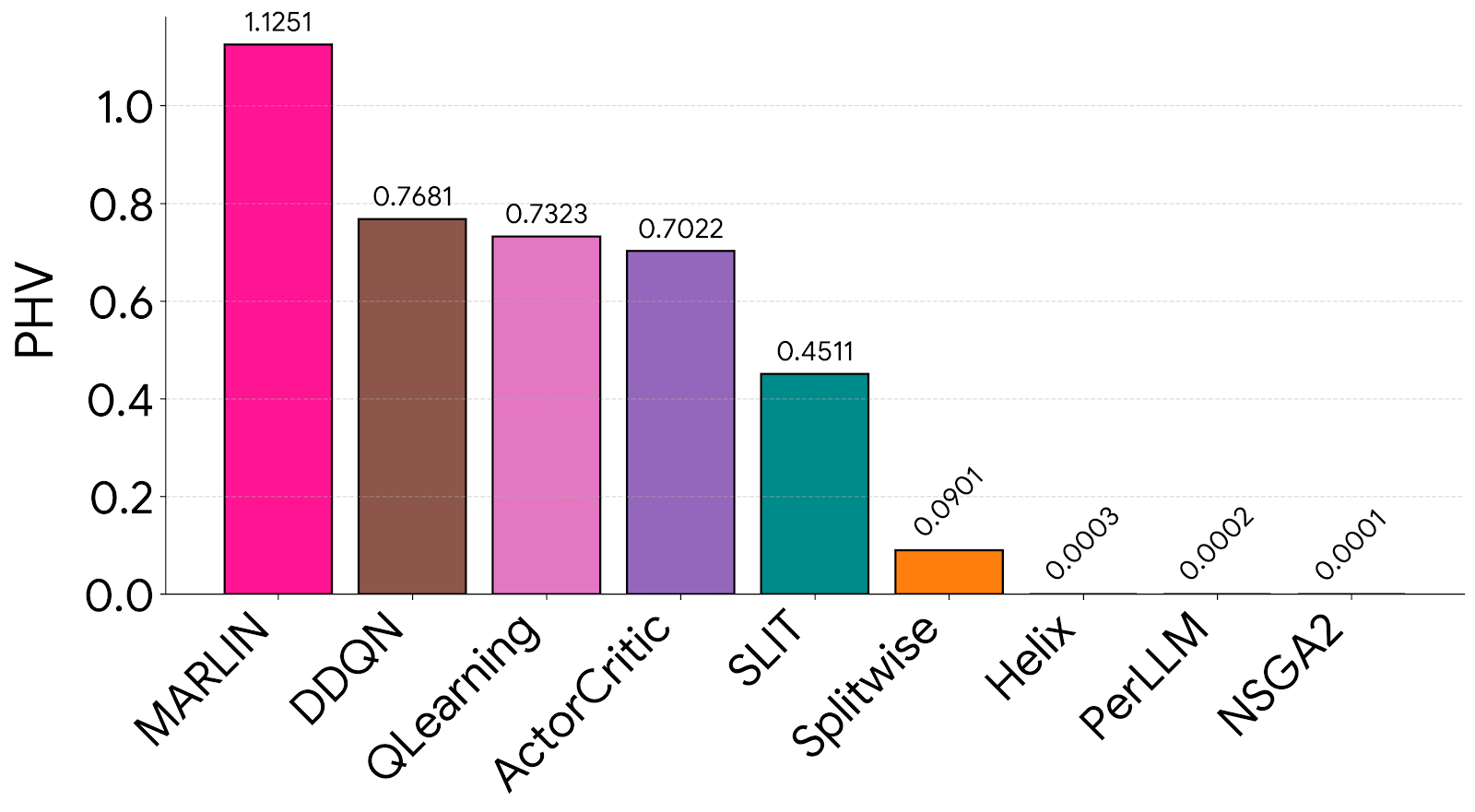}
  \caption{Comparison of the PHV values across LLM inference scheduling frameworks.}
  \Description{Vertical bar chart comparing Pareto hypervolume (PHV) values for nine LLM inference scheduling frameworks. The horizontal axis lists frameworks left to right in descending PHV order. The vertical axis shows PHV from 0 to 1.2. MARLIN (magenta bar) achieves the highest PHV of 1.1251. DDQN (brown) follows at 0.7681, QLearning (pink) at 0.7323, and ActorCritic (purple) at 0.7022. SLIT (teal) is notably lower at 0.4511. Splitwise (orange) drops sharply to 0.0901. The remaining three heuristic frameworks---Helix, PerLLM, and NSGA2---have near-zero PHV values of 0.0003, 0.0002, and 0.0001 respectively, with bars barely visible above the axis.}
  \label{fig4:PHV_Frameworks}
\end{figure}
\begin{figure*}
  \centering
  \includegraphics[width=\linewidth]{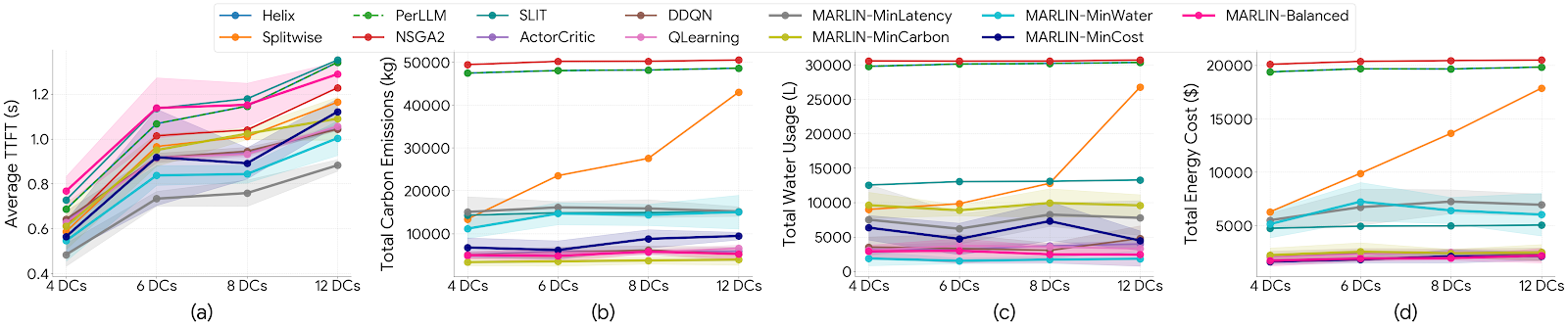}
  \caption{Comparison of (a) TTFT, (b) carbon emissions, (c) water usage, and (d) energy costs across LLM inference scheduling frameworks as the number of datacenters changes between 4 and 12; shaded regions represent the confidence intervals.}
  \Description{Four-panel line chart showing scalability of 13 LLM inference scheduling frameworks as the number of datacenters increases from 4 to 6, 8, and 12. Lines with circular markers are shown for each framework; shaded bands around RL-based lines indicate 95 percent confidence intervals. The legend matches that of Figure 3. Panel (a) shows average TTFT in seconds. All frameworks increase from roughly 0.4 to 0.6 seconds at 4 datacenters to 0.8 to 1.3 seconds at 12 datacenters. MARLIN-MinLatency maintains the lowest trajectory; MARLIN-Balanced rises more steeply but remains competitive. Panel (b) shows total carbon emissions in kilograms. Helix, PerLLM, and NSGA2 remain flat near 47000 to 50000 kg across all datacenter counts. Splitwise increases sharply from roughly 15000 kg at 4 datacenters to over 42000 kg at 12 datacenters. MARLIN-MinCarbon, MARLIN-MinCost, MARLIN-Balanced, and SLIT remain low and roughly flat between 5000 and 15000 kg. Panel (c) shows total water usage in liters. Helix, PerLLM, and NSGA2 stay near 30000 L. Splitwise rises from about 8000 L at 4 datacenters to over 27000 L at 12. MARLIN variants hold low values, with MARLIN-MinWater staying near 1000 to 2000 L. Panel (d) shows total energy cost in dollars. Helix, PerLLM, and NSGA2 remain near 19000 to 21000 dollars. Splitwise rises from roughly 5000 dollars at 4 datacenters to over 18000 dollars at 12. MARLIN-Balanced stays the lowest across all datacenter counts, near 1500 to 2000 dollars.}
  \label{fig5:scalability_results}
\end{figure*}
\section{Experimental Results}
To empirically validate our framework, MARLIN, we compare it against eight state-of-the-art workload scheduling algorithms. These can be divided into five heuristic-based approaches (Helix \cite{10.1145/3669940.3707215}, Splitwise \cite{10609649}, NSGA-II \cite{996017}, PerLLM \cite{yang2024perllmpersonalizedinferencescheduling}, and SLIT \cite{10.1145/3716368.3735301}), and three RL methods (QLearning \cite{10791320}, DDQN \cite{10.1145/3773274.3774691}, and ActorCritic \cite{11175406}). Datacenters are distributed globally with requests originating uniformly across all regions. Each datacenter has 1000 compute nodes across 6 uniformly distributed node types, each containing 2, 4, or 8 NVIDIA A100 or H100 GPUs for LLM inference on Xeon-based server nodes. We developed and validated a Python simulator implementing the models from Section 3, as well as the workload predictor and MARLIN algorithmic framework. Data for computational energy and LLM runtime is obtained from profiling on real server nodes and integrated into the simulator.

We consider five schemes to be run through our framework, which consist of four plans dominated by one agent (MARLIN-MinCarbon, -MinWater, -MinCost, -MinLatency) and one with equal weights (MARLIN-Balanced). Our baseline experimental setup spans 24 hours with the maximum datacenter utilization rate in an epoch at 95\% across 8 datacenters. The upcoming epoch’s workload is forecast (a 15-minute window) in the previous epoch.

The learning rate hyperparameter in the MARLIN framework was 0.0003 for the actor and 0.001 for the critic networks. We used a gamma of 0.95 and a tau of 0.005 for the agents. The actor network has 128 hidden dimensions, while the critic was increased to 256 to capture the complexity of the design space. The replay buffer size was 20,000 samples for the current epoch and 5,000 samples for previous epochs, with a 30/70 sampling split. For the SGD step, we used a learning rate of 0.05 with 5 steps. To determine the strength and use of the veto, the required threshold was 150 capital, with a 0.5 pull towards the agent’s own plan when used.
\subsection{State-of-the-art Comparison}
Fig. \ref{fig3:baseline_results} shows MARLIN's performance compared with state-of-the-art frameworks across the TTFT, Carbon, Water, and Energy Cost metrics, with confidence bounds for all RL frameworks. The MARLIN variants outperform all comparison works across these metrics, even considering confidence bounds. For the single-metric schemes, MARLIN-MinLatency reduces TTFT by 18.67\% over QLearning, MARLIN-MinCarbon reduces carbon emissions 33.61\% more than QLearning, MARLIN-MinWater reduces water 43.88\% more than DDQN, and MARLIN-MinCost reduces costs 11.72\% more than ActorCritic. MARLIN-Balanced outperforms the heuristic approaches across all metrics except TTFT and outperforms the RL baselines on 2 of 4 metrics; it is at most 23.61\% slower in TTFT and 2.77\% higher in carbon emissions than QLearning, while achieving at least an 18.66\% reduction in water (vs. DDQN) and 18.94\% cost reduction (vs. ActorCritic). Compared to the only other explicitly sustainable framework, SLIT, MARLIN-Balanced achieves a 2.29\% reduction in TTFT, 61.07\% in carbon, 81.11\% in water, and 60.90\% in total costs — demonstrating that MARLIN succeeds as a sustainable framework where SLIT failed.

In Fig. \ref{fig4:PHV_Frameworks}, we summarize the Pareto hypervolume (PHV) values across the different frameworks. The PHV of a Pareto solution front (the front formed by the non-dominated solutions) is a metric that measures the hypervolume of the solution space dominated by the set of solutions. This is an effective measure of the quality of diverse solution sets generated by a multi-objective optimization framework. A higher PHV indicates a more diverse and higher quality solution set. 

For comparison works that produce only a single point, the PHV is calculated as the geometric volume of the four-dimensional hyperrectangle defined by that point. We set a reference point for the worst observed metrics. The population of points along the Pareto front was 1 for heuristic frameworks. For the RL frameworks, we archived the best points generated during the search along the confidence interval providing between 10 and 15 points. MARLIN’s Pareto front contained 40 points, well exceeding the value of the other frameworks. We observe a significant difference between the heuristic-based and the RL-based approaches. The best-performing heuristic framework is SLIT, which covers only 40.09\% of the hypervolume covered by MARLIN. The RL frameworks perform much better, but still only cover at most 68.27\% of the volume that MARLIN covers. These results show that MARLIN is able to search the complex solution space much more effectively than the other frameworks, which allows it to provide higher quality solutions.

\subsection{Scalability Analysis}

To observe how the frameworks behave as the problem size scales, we vary the number of datacenters from 4 to 12. The results of this experiment are presented in Fig. \ref{fig5:scalability_results}. 

It can be observed that the heuristic approaches were unable to utilize the new datacenters effectively and to navigate the larger search space, as datacenter counts increased. The RL-based approaches have more success in utilizing the new datacenters and their different carbon and water intensities. 

MARLIN variants dominate, with MARLIN-MinCarbon being an outlier as it outperformed MARLIN-MinCost on cost with 12 datacenters. This performance by the carbon agent highlights a limitation of the single-metric variants. Low carbon sources can be low cost, but the signal in the carbon case is stronger than the cost case. MARLIN-MinCarbon finds this strong signal early and takes advantage of low carbon and cheap electricity for longer than MARLIN-MinCost, leading to the discrepancy. 

Crucially, the MARLIN-Balanced approach can achieve Pareto-optimal solutions as the search space grows. This is significant, MARLIN-Balanced can reduce water usage by an average of 16.4\% compared to smaller search spaces. MARLIN-Balanced leverages the unique sustainability fingerprint of each datacenter it has access to. In this experiment, we observe the same trend between MARLIN-Balanced and SLIT. In 12 datacenters, MARLIN-Balanced reduces latency by 4.66\%, carbon emissions by 65.22\%, water usage by 81.7\%, and cost by 56.59\%. MARLIN-Balanced provides a more balanced solution instead of the best across all metrics. 
\subsection{MARLIN Framework Ablation Study}

\begin{figure}
  \centering
  \includegraphics[width=\linewidth]{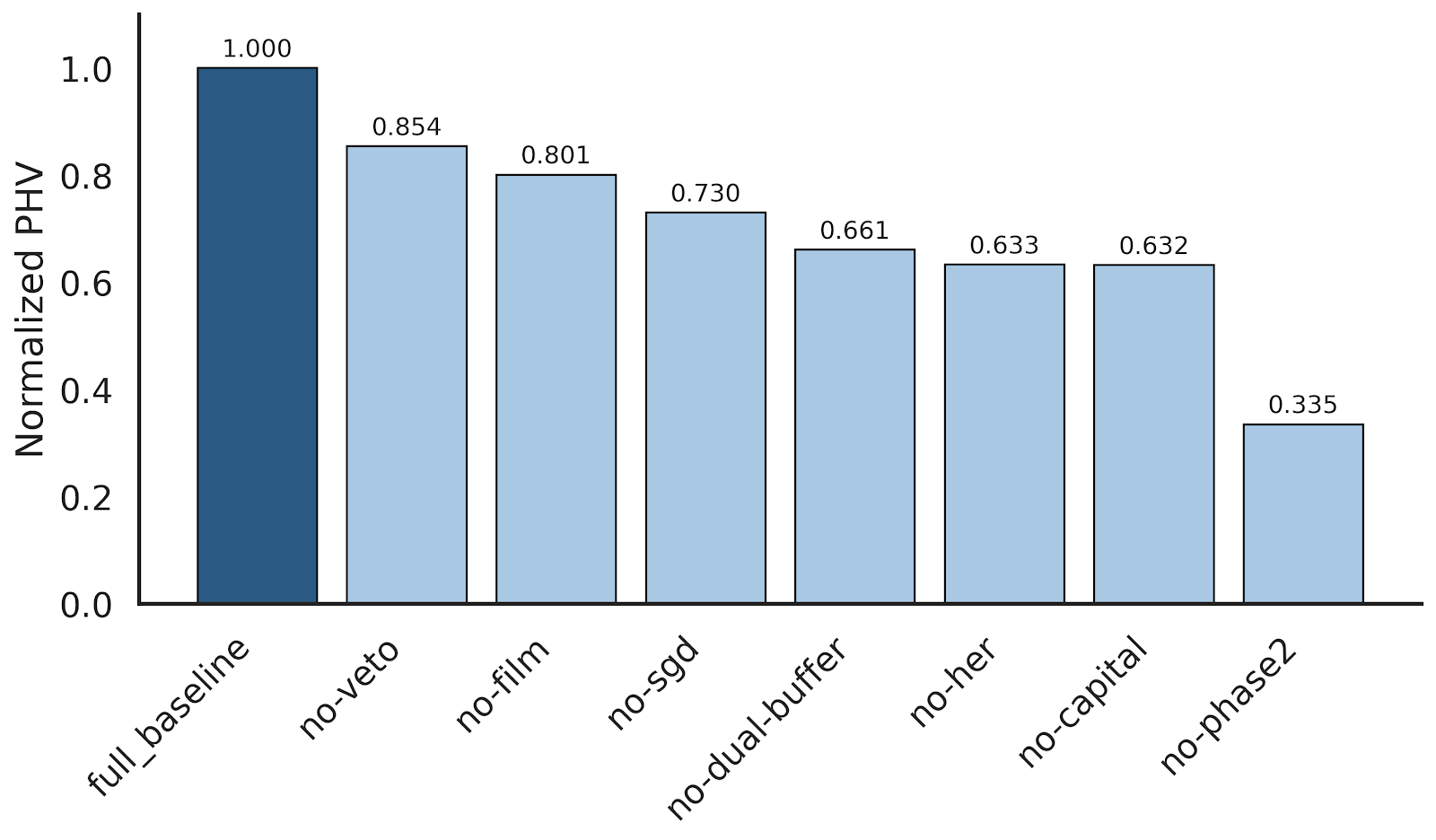}
  \caption{Normalized PHV of MARLIN framework (full\_baseline) compared to various ablations.}
  \Description{Vertical bar chart showing a MARLIN ablation study using normalized PHV on the vertical axis from 0 to 1.0. Eight configurations are shown left to right in descending order of normalized PHV. The full baseline (dark navy bar) achieves a normalized PHV of 1.000. The remaining seven ablations are shown as light blue bars: no-veto at 0.854, no-film at 0.801, no-sgd at 0.730, no-dual-buffer at 0.661, no-her at 0.633, no-capital at 0.632, and no-phase2 at 0.335. The largest drop is caused by removing phase 2, which reduces PHV by 66.5 percent relative to the full baseline. Removing only the veto causes the smallest drop of 14.6 percent.}
  \label{fig6:Ablation_Study}
\end{figure}

We analyzed the contribution of each component of the MARLIN framework by selectively removing them in an ablation study. The experiments use baseline parameters. The results of the ablation study are shown in Fig. \ref{fig6:Ablation_Study}.

The plot shows that the PHV improvement of the MARLIN (full\_baseline) against the best-performing ablation is at least 14.6\%. We can observe that all parts of the MARLIN framework are important towards improving the quality of the generated solutions. For instance, without the least impacting component Veto, the PHV of MARLIN will decrease 14.9\%. Without the blending process in phase 2, the PHV will decrease 66.5\%.
\section{Conclusion}

In this work, we presented a novel framework for serving LLM inference requests across geo-distributed cloud datacenters, called MARLIN. We empirically demonstrated that MARLIN can optimize the TTFT of LLM inference requests while also optimizing sustainability metrics such as carbon emissions and water usage. Solutions generated by the MARLIN framework outperformed other state-of-the-art frameworks and produced a balanced scheduling plan when configured with equal metric weights. This was observed in our baseline experiments, where MARLIN, decreased TTFT by 18\%, carbon emissions by 33\%, water usage by 43\%, and costs by 11\%. MARLIN is designed to operate as a meta scheduler above containers that host LLMs, such as Kubernetes or vLLM. This allows MARLIN to be used by datacenter managers to provide a range of LLM inference services and enable sustainable, multi-objective decision-making. Our future work will explore integrating embodied carbon to optimize life-cycle carbon emissions for LLM serving.   
\begin{acks}
This research was made possible with support from HPE and grants from the National Science Foundation (CCF-2324514, CNS-2132385).
\end{acks}

\bibliographystyle{ieeetr}
\bibliography{MARLIN}

\end{document}